\documentstyle[12pt]{article}

\advance\hoffset by -7mm
\makeatletter
\@addtoreset{equation}{section}
\makeatother

\renewcommand{\title}[1]{\null\vspace{25mm}

\noindent{\Large{\bf #1}}\vspace{10mm}

\noindent {\large By }}
\newcommand{\authors}[1]{\noindent{\large #1}\vspace{3mm}

}
\newcommand{\address}[1]{\noindent #1\vspace{5mm}

}
\renewcommand{\abstract}[1]{\vspace{19mm}

\noindent{\small{\em Abstract.} #1}\vspace{2mm}

}

\newcommand{\F}{\noindent}

\newcommand{\q}{\qquad}

\newcommand{\beq}{\begin{eqnarray}}
\newcommand{\ene}{\end{eqnarray}}

\newcommand{\HH}{{\cal H}}
\newcommand{\UU}{{\cal U}}
\newcommand{\OO}{{\cal O}}
\newcommand{\la}{{\lambda}}

\newcommand{\Ltnn}{{L^2(R^{3n})}}

\newcommand{\ve}{\vert}

\begin{document}

\title{Theory of Local Times\\[2mm]
 II. Another formulation and examples}

\normalsize

\authors{Hitoshi Kitada}

\address{Department of Mathematical Sciences, University of Tokyo\\
Komaba, Meguro, Tokyo 153, Japan}

\abstract{
The model of a stationary universe and the notion of local times presented
in \cite{K} are reviewed with some alternative formulation of the consistent
 unification of the Riemannian and Euclidean geometries of general
relativity and quantum mechanics. The method of unification adopted in the
 present paper is by constructing a vector bundle $X\times R^6$ or $X\times
 R^4$ with $X$ being the observer's reference frame and $R^6$ or $R^4$
being the unobservable inner space(-time) within each observer's local
 system. Some applications of our theory to two concrete examples
 of human size and of cosmological size are
 discussed, as well as the uncertainty of time in our context is
calculated.}

\section{Introduction}

As stated in Section 2 of \cite{K}, it seems that there are fundamental
 difficulties in the presently existing theories trying to unify the
relativity and quantum mechanics.

We review quickly and briefly some fundamental difficulties of these
theories that try to unify the quantum and relativity theories.

\F
{\bf The difficulties of the existing theories:}

\F
1) The quantum field theories---unification of special relativity and
quantum theory

In the trials in this direction, there is the difficulty of the
 divergence, or the problem of renormalization. According to the
 Euclidean method in the axiomatic quantum field theory that is a
 mathematically rigorous investigation in this direction, the
 following is known. We denote by $\nu$ the spacetime dimension:

a) For $\nu\ge 5$, there are no non-trivial models.

b) For $\nu=2, 3$, there are non-trivial models.

c) For $\nu=4$,
there are no non-trivial models,
if one use the lattice approximation with some additional assumptions
 on the renormalizability of the theories.

For the details, see, e.g., Streater's paper in \cite{3},
 Fr\"ohlich \cite{F}.

\F
2) The quantum gravity---unification of general relativity and
 quantum theory

In the quantum mechanics, time occupies a special position as
 in the Newtonian mechanics. The Newtonian time is a concept which
 is incompatible with diffeomorphism-invariance.
Here is a central
 problem. Namely if the general relativistic quantum mechanics
 is completed, then the (proper) time should be defined as an invariant
 quantity with respect to the diffeomorphisms group of spacetime.
 But in the quantum theories, time remains an absolute notion in the
 Newtonian sense. There are various trials in treating this difficulty.

Among the trials where the problem is regarded as the quantization of
 general relativity, there are two directions of trials. One is to
 identify time before quantization, and another is to identify time
 after quantization.

There are also trials without assuming the notion of time.

For details, see, e.g., Isham \cite{I}. See also Ashtekar and
 Stachel \cite{2}.

In \cite{K}, we presented a formulation, where the notion of
 local times eliminates several fundamental difficulties mentioned above.

The purpose of the present part II is to give some simpler exposition
 of the consistent unification of quantum and relativity theories with
 clarifying the role of the observer's systems, and their relation with
 the inner spacetime within each local system. In short, the unification
 in this part II is done by orthogonalizing the observer's coordinates
 (= Riemannian spacetime) to the Euclidean spaces inside each local
 system. This presentation is a realization of the remark stated in
 Section 6 of \cite{K}, and will be done in Sections 2--4 below with
 some repetitions of the axioms and definitions stated in \cite{K}.

As a concrete exposition of our theory, we will give in Section 5 an
 explanation of the experiment of the interference of one neutron in
 the uniform gravitational field, done by Collela et al. \cite{C}.
 We will also explain the Hubble's redshift in Section 6, which will
 be done in quite the same way as in the general relativity theory.
 In the final section 7, we will calculate the order of the
 uncertainty of time which is predicted by our theory.

\section{Local times}

\F
{\bf Fundamental 3 axioms:}

\normalsize

Let $\HH$ be a separable Hilbert space, and set
\beq
\UU=\{\phi\}=\bigoplus_{n=0}^\infty \left(\bigoplus_{\ell=0}^\infty
\HH^n  \right) \q (\HH^n=\underbrace{\HH\otimes\cdots\otimes\HH}_{n\
\mbox{\scriptsize{factors}}}).
\ene
$\UU$ is called a (separable) Hilbert space of possible universes.

Let $\OO=\{ A\}$ be the totality of the selfadjoint operators $A$
 in $\UU$ of the form
 $A\phi=(A_{n\ell}\phi_{n\ell})$ for
$\phi=(\phi_{n\ell})\in\UU$ in the domain of the operator $A$.

\F
{\bf Axiom 1.}\
There exist
 a selfadjoint operator $H\in\OO$ in $\UU$
such that for some $\phi\in \UU-\{0\}$ and $\la\in R^1$
\beq
H\phi=\la \phi
\ene
in the following sense:
There exists an infinite matrix $(\la_{n\ell})$ of real numbers such that
$H_{n\ell}\phi_{n\ell}=\la_{n\ell}\phi_{n\ell}$ for each $n\ge1$,
 $\ell\ge 0$ and $\la_{n\ell_n}\to\la$ as $n\to\infty$ along any $\ell_n$
 such that $F_n^{\ell_n}\subset F_{n+1}^{\ell_{n+1}}$. Here $F_n$ is
 a finite subset of ${\mbox{\bf N}}=\{1,2,\cdots\}$ with $\sharp(F_n)=$
 (the number of elements in $F_n)=n$ and
$\{ F_n^\ell\}_{\ell=0}^\infty$ is the totality of such $F_n$.

$H$ is an infinite matrix $(H_{n\ell})$ of selfadjoint operators
 $H_{n\ell}$ in $\HH^n$. Axiom 1 asserts that this matrix converges
 in the sense of (2.2).

\F
{\bf Axiom 2.}\
Let $n\ge 1$ and $F_{n+1}$ be a finite subset of
${\mbox{\bf N}}=\{1,2,\cdots\}$ with
$\sharp(F_{n+1})=n+1$. Then for any  $j\in F_{n+1}$, there exist
 selfadjoint operators $X_j =(X_{j 1},X_{j2},X_{j3})$,
 $P_j =(P_{j1},P_{j2},P_{j3})$ in
$\HH^n$ and constants $m_j>0$ such that
\beq
[X_{j\ell},X_{k m}]=0,\q [P_{j\ell},P_{k m}]=0,\q [X_{j\ell},P_{k m}]=
i\delta_{jk}\delta_{\ell m},
\ene
\beq
\sum_{j\in F_{n+1}} m_j X_j=0,\q
\sum_{j\in F_{n+1}} P_j=0.
\ene

By the Stone-von Neumann theorem Axiom 2 also specifies the space dimension
(see  Abraham-Marsden \cite{1}, p.452). We identify $\HH^n$
with $\Ltnn$ in the following.

What we want to mean by $H_{n\ell}$ $(n,\ell\ge 0)$ in Axiom 1 is the $N=n+1$
body Hamiltonian in the usual quantum mechanics. For  the local Hamiltonian
$H_{n\ell}$ we thus make the following postulate.

\F
{\bf Axiom 3.}\ Let $n\ge0$ and $F_N$ $(N=n+1)$ be a finite subset of
${\mbox{\bf N}}=\{1,2,\cdots\}$ with $\sharp(F_N)=N$.
Let $\{ F_N^\ell\}_{\ell=0}^\infty$ be
the totality of such $F_N$. Then the Hamiltonians
$H_{n\ell}$ $(\ell\ge0)$ are of the form
\beq
H_{n\ell}=H_{n\ell0}+V_{n\ell},\q V_{n\ell}=\sum_{\alpha=(i,j), 1
\le i<j<\infty, i,j\in F_N^\ell} V_\alpha(x_\alpha)
\ene
on $C_0^\infty(R^{3n})$, where $x_\alpha=x_i-x_j$ with $x_i$ being the
position vector of the $i$-th particle, and $V_\alpha(x_\alpha)$ is a
real-valued measurable function of $x_\alpha\in R^3$ which is
$H_{n\ell0}$-bounded with $H_{n\ell0}$-bound of $V_{n\ell}$ less than 1.
$H_{n\ell0}=H_{(N-1)\ell0}$ is the free Hamiltonian of the
$N$-particle system.  The concrete form is expressed as in  \cite{15},
 (1.4), if one use clustered Jacobi coordinates.

This axiom implies that $H_{n\ell}=H_{(N-1)\ell}$ is uniquely extended to a
selfadjoint operator in $\HH^n=\HH^{N-1} =L^2(R^{3(N-1)})$ by the
Kato-Rellich theorem.

\vskip 18pt

\F
{\bf A theorem in many-body scattering theory:}
\normalsize

For the $N$-body Hamiltonian $H_{N-1}=H_{n\ell}$ $(N=n+1)$ the following
Theorem 1 is known \cite{6}  to hold under suitable assumptions
 (e.g., Assumptions 1 and 2 in \cite{15}).

We here follow the notation and conventions in \cite{15} for the $N$-body
quantum systems. In particular $H_b=H_{(N-1)b}=H_{N-1}-I_b=H^b_{n\ell}
+T_{n\ell b}=H^b+T_b$ is the truncated Hamiltonian for the cluster
decomposition $1\le\ve b\ve\le N$, and  $P_b^M$ is the $M$-dimensional
partial projection of the eigenprojection $P_b=P_{H^b}$ associated with
the subsystem $H^b$, i.e., $P_b$ is the orthogonal projection in
$\HH^b=L^2(R^{3(N-\vert b\vert )})$
onto the eigenspace of $H^b$. $q_b$ is the velocity conjugate to the
intercluster coordinates $x_b$. We define for
   a $k$-dimensional multi-index    $M=(M_1,\cdots,M_k)$, $ M_j \ge 1$,
\beq
   {\hat P} ^M_{k}= \left(I-\sum_{\vert b \vert = k}P_b^{M_k}\right) \cdots
   \left(I-\sum_{\vert d \vert = 2}
   P_d^{M_{2}}\right)
   (I-P^{M_1}),\\
\nonumber
      k=1,\cdots ,N-1,
\ene
where $P^{M_1}=P_a^{M_1}$ with $\ve a\ve=1$, and for
a $\ve b\ve$-dimensional
multi-index $M_{b} = (M_1, \cdots ,$ $ M_{\vert b \vert -1},
M_{\vert b \vert})$  $ = ({\hat M}_{b}, M_{\vert b \vert})$
\beq
   {\tilde P}_{b}^{M_{b}}=P_b^{M_{\vert b \vert}}{\hat P}_{\vert b \vert
   -1}^{{\hat M}_{b}}, \q 2\le\ve b\ve\le N.
\ene
It is clear that
\beq
\sum_{2\le\ve b\ve\le N} {\tilde P}^{M_{b}}_{b} = I - P^{M_1},
\ene
provided that the component $M_k$ of $M_{b}$ depends only on the number
$k$ but  not on  $b$.  In the following we use such $M_{b}$'s only.
 Under these circumstances, the following is known to hold.

\F
{\bf Theorem 1}
(\cite{6} Enss). Let Assumptions 1 and 2 in {\rm{\cite{15}}} be
 satisfied.  Let $f \in \HH^{N-1}$.  Then there  are a sequence $t_m \to
\pm\infty$ (as $m \to \pm\infty)$ and a sequence  $M^m_{b}$ of
multi-indices whose components all tend to
$\infty$ as $m \to \pm \infty$ such
that for all cluster decompositions $b$, $2\le\ve b\ve\le N$,  $\psi \in
C_0^\infty (R^1)$, and $ \varphi \in C_0^\infty (R^{3(\vert b \vert -1)})$
\beq
   \left\Vert \frac {\vert x^b \vert ^2}{t^2_m} {\tilde P}_{b}^{M^m_{b}}
   e^{-it_m H_{N-1}} f \right\Vert \to 0,
\ene
\beq
\Vert \{ \psi (H_{N-1}) - \psi (H_b) \} {\tilde
   P}_{b}^{M^m_{b}} e^{-it_m H_{N-1}} f \Vert \to 0,
\ene
\beq
   \Vert \{ \varphi (x_b/t_m) - \varphi (q_b) \} {\tilde P}_{b}^{M^m_{b}}
   e^{-it_m H_{N-1}} f \Vert \to 0
\ene
as $m \to \pm\infty$.

\F
{\bf Definition of local times:}
\normalsize

\F
{\bf Definition 1.}
\begin{itemize}
\item
 Let $\phi=(\phi_{n\ell})$ with
$\phi_{n\ell}=\phi_{n\ell} (x_1,\cdots,x_n)\in \Ltnn$
be the universe in Axiom 1.
\item
We define $\HH_{n\ell}$
as the sub-Hilbert space of $\HH^n$ generated by the functions
$\phi_{mk}(x^{(\ell)},y)$ of $x^{(\ell)}\in R^{3n}$ with regarding $y\in
R^{3(m-n)}$ as a parameter, where $m\ge n$,
$F_{n+1}^\ell\subset F_{m+1}^{k}$,
 and $x^{(\ell)}$ is the (relative) coordinates of $n+1$ particles in
$F_{n+1}^\ell$.
\item
$\HH_{n\ell}$ is called a {\bf local universe} of  $\phi$.
\item
$\HH_{n\ell}$ is said to be non-trivial if $(I-P_{H_{n\ell}})\HH_{n\ell}\ne
\{0\}$.
\end{itemize}

The total universe $\phi$ is a single element in \ $\UU$. The local universe
$\HH_{n\ell}$ may be richer and may have elements more than one. This is
because we consider the subsystems of the
universe consisting of a finite number of particles. These subsystems
 receive the influence from the other particles of infinite number
outside the subsystems, and may vary to constitute a non-trivial subspace
$\HH_{n\ell}$.

\F
{\bf Definition 2.}
\begin{itemize}
\item
The restriction of $H$ to $\HH_{n\ell}$ is also denoted by the same
notation  $H_{n\ell}$ as the $(n,\ell)$-th component of $H$.
\item
We call the pair
$(H_{n\ell},\HH_{n\ell})$  a local system.
\item
The unitary group
$e^{-itH_{n\ell}}$ $(t\in R^1)$ on $\HH_{n\ell}$ is called  the {\bf proper
clock} of the local system $(H_{n\ell},\HH_{n\ell})$, if $\HH_{n\ell}$ is
non-trivial: $(I-P_{H_{n\ell}})\HH_{n\ell}\ne \{0\}$.
\item
Note that the clock is
defined only for $N=n+1\ge 2$, since $H_{0\ell}=0$.
\item
 The universe $\phi$ is
called {\bf rich} if $\HH_{n\ell}$ equals
$\HH^n=L^2(R^{3n})$ for all $n\ge 1$, $\ell\ge0$. For a rich universe $\phi$,
$H_{n\ell}$ equals the $(n,\ell)$-th component of $H$.
\end{itemize}

The formula (2.11) indicates that $t_m$ is asymptotically equal to $\pm\ve
x_b\ve/\ve q_b\ve$ as $m\to\pm\infty$, independently of the choice of cluster
decompositions $b$. This is precisely the actual procedure of measuring the
time $t_m$ in the mechanics. The implication of this theorem is therefore
interpreted as follows: If one \lq measures' the time of a state $f\in
(I-P_{H_{(N-1)\ell}})\HH_{(N-1) \ell}-\{0\}$ in the local system
$(H_{(N-1)\ell},\HH_{(N-1)\ell})$ by the associated proper clock
$e^{-itH_{(N-1)\ell}}f$, namely if one measures the quotient
$\pm\ve x_b\ve/\ve q_b\ve$ of the scattered particles which are regarded as
moving almost in a steady velocity, then that time is asymptotically equal to
the parameter $t_m$ in the exponent of $e^{-it_m H_{(N-1)\ell}}f$  as
$m\to\pm\infty$. In this sense $t_m$ is interpreted as the
 {\bf quantum mechanical proper time} of
 the local system $(H_{n\ell},\HH_{n\ell})=(H_{(N-1)\ell},
\HH_{(N-1)\ell})$, if $(I-P_{H_{(N-1)\ell}})\HH_{(N-1)\ell}\ne \{0\}$.

\F
{\bf Definition 3.}
\begin{itemize}
\item
The parameter $t$ in the exponent of the proper clock $e^{-itH_{n\ell}}=
e^{-itH_{(N-1)\ell}}$ of the local system $(H_{n\ell},\HH_{n\ell})$
is called the (quantum mechanical) {\bf proper time}
or {\bf local time} of the local system $(H_{n\ell}, \HH_{n\ell})$,
if $(I-P_{H_{n\ell}})\HH_{n\ell}\ne \{0\}$.
\item
This time $t$ is denoted by $t_{(H_{n\ell},\HH_{n\ell})}$
indicating the local system under consideration.
\end{itemize}

This definition is the one reverse to the usual definition of the motion or
dynamics of the $N$-body quantum systems, where the time $t$ is given
 {\it a priori} and the motion of the particles is defined by
 $e^{-itH_{(N-1)\ell}}f$
for a given initial state $f$ of the system.

{\bf Time} is thus defined only for the local systems
$(H_{n\ell},\HH_{n\ell})$  and is determined by the associated proper
 clock $e^{-itH_{n\ell}}$. Therefore
 there are infinitely many times $t=t_{(H_{n\ell},\HH_{n\ell})}$ each
of which  is proper to the local system $(H_{n\ell},\HH_{n\ell})$. In this
 sense time is a local notion. There is no time for the total universe
$\phi$ in Axiom 1, which is a (stationary) bound state for the total
Hamiltonian $H$.

\F
{\bf Uncertainty relation among time, position and momentum:}

This local time is an approximate one in a double sense.

\begin{itemize}
\item
 First $t_m$ is only
{\it asymptotically} equal to $\pm\ve x_b\ve/\ve q_b\ve$ as $m\to\pm\infty$.

This fact explains the so-called principle of uncertainty in our context.
In the usual explanation, the position $x_b$ and the velocity $q_b$ or the
momentum $p_b$ cannot be determined in equal accuracy.
According to our theory,
 this is rephrased as follows: The time $t$ cannot be determined accurately,
even if $x_b$ and $q_b$ could be determined precisely.
It is only determined in
 some {\it mean} sense as in (2.11).
{}From the usual uncertainty relation $\Delta x\cdot\Delta p\ge \hbar/2$
 and $p=mq$ follows $\Delta x\cdot\Delta q\ge \hbar/(2m)$, which
 indicates that the uncertainty of time is proportional to $m^{-1}$.
 This explanation resolves the difficulty of the uncertainty between
 time and energy when one considers the time as an operator.
(We will give a concrete explanation in Section 7.)

\item
Second the local  Hamiltonian $H_{n\ell}$ is not the total Hamiltonian $H$.
Or rather, the time arises from this approximation of $H$ by $H_{n\ell}$.

This approximation may make $\HH_{n\ell}$ non-trivial, and the clock
$e^{-itH_{n\ell}}$ can be defined as in Definition 2 owing to
$(I-P_{H_{n\ell}})\HH_{n\ell}\ne \{0\}$.

On the contrary the total universe
$\phi$ has no associated clock and time, since $(I-P_H)\phi=0$.
\end{itemize}

\F
{\bf Mutual independency of local systems:}

Our theory of local times further implies in particular that local systems
$(H_{n\ell},\HH_{n\ell})$  are {\it mutually independent}.

{\it Indication of Proof:}\
As a typical example, let us consider the case
$F_{N^\prime}^{\ell^\prime}\subset F_N^\ell$ with $N^\prime< N$.
 In this case, $H_{(N^\prime-1)\ell^\prime}$ is a subsystem Hamiltonian
 of $H_{(N-1)\ell}$. However the correspondent times
 $t_{N^\prime\ell^\prime}$ and $t_{N\ell}$ are
measured mutually independently  as in Theorem 1-(2.11).
\vskip-2mm

The problem that may arise in this case is with the
 common variable $x^{(\ell')}$. But as $\HH_{(N'-1)\ell'}$
 and $\HH_{(N-1)\ell}$ indicate, these spaces have different
 $(\ell',N')$ and $(\ell,N)$. Thus by (2.1) before Definition 1, these spaces
 $\HH_{(N'-1)\ell'}$ and $\HH_{(N-1)\ell}$
are the subspaces of the $\ell'$-th $\HH^{(N'-1)}$ and $\ell$-th
 $\HH^{(N-1)}$ of the formula (2.1), hence are mutually independent
 Hilbert spaces. This implies that the $L^2$-representations of
 these spaces are also mutually independent. Therefore the correspondent
 clocks and local times are also mutually independent.

\F
{\bf Interpretation of usual quantum mechanics:}

We have defined the (local) time $t=t_{(H_{n\ell},\HH_{n\ell})}$ for each
local  system $(H_{n\ell},\HH_{n\ell})$. This time $t$ satisfies Theorem
1-(2.11). If one regards the time $t$ as a {\it given} quantity, this fact is
interpreted as follows: In each local system $(H_{n\ell},\HH_{n\ell})$, the
physics  follows the quantum mechanics, i.e., follows the Schr\" odinger
equation.

\F
{\bf Toward the relativity:}

Our definition of times is consistent with the theory of (general)
 relativity of Einstein. Our (quantum mechanical) proper time
 of the local system $(H_{n\ell}, \HH_{n\ell})$ can be regarded
 as the quantum mechanical correspondent to
the classical proper time in the theory of relativity.

\section{Relativity}

For the relative motions of the {\it centers} of mass of local systems, we
postulate the principle of (general) relativity and the principle of
equivalence as in Einstein \cite{7}.

\F
{\bf Fundamental assumptions on `observable' and `unobservable':}

What should be stated first on our introduction of relativity is that
we make the following fundamental assumptions:

\begin{itemize}
\item
{\bf
Only} the relative classical motions of the {\bf centers} of mass of local
systems are {\bf observable} in our theory.
\item
The {\bf internal} quantum
mechanical motion within each local system is independent of {\bf
observation},  at the present stage of our theory. In
this sense, the internal quantum mechanical motion within a local system is
{\bf unobservable}.
\item
We postulate Axiom 6 in the next section 4,
which gives a principle of {\bf deduction} of the {\bf internal}
quantum mechanical motion within each local system from classical
observations of its {\bf sub} local systems, through certain relativistic
 considerations.
\end{itemize}

\F
{\bf Identification of local time with the relativistic proper time:}

Fix one observer's local system $L_O=(H_{n\ell},\HH_{n\ell})$.
 Then the classical world observable by $L_O$ is observed within
 the 4-dimensional Riemann space $X$, whose time is assumed
 to coincide with the local time $t$ of that local system $L_O$
 at the center of mass of $L_O$, and whose origin of the space
 coordinates is assumed to be equal to the $L_O$'s center of mass.

The unobservable inner space associated with the local system $L_O$
 is the Euclidean space $R^6$ consisting of the points $(x,p)$,
 where $x$ is the configuration and $p$ is the momentum conjugate
 to $x$. The local time $t$ is defined as the parameter $t$ of the
 exponent of the proper clock $e^{-itH_{n\ell}}$ of that local system
 $L_O$ (Definitions 1--3). Then $R^6$ can be regarded as $R^4$ with
 coordinates $(t,x)=(t_{(H_{n\ell},\HH_{n\ell})},x_{(H_{n\ell},
\HH_{n\ell})})$.

In this way we have two different frameworks or coordinate systems
 for the observable classical world and unobservable quantum world,
 respectively. These coordinate systems are `orthogonal' in the
 vector bundle{\footnote[2]{Here $X\times R^6$ is a trivial vector
 bundle with base space $X$ and fibre $R^6$. Thus this vector bundle
 can be identified with the direct product of a Riemannian manifold
 $X$ and a Euclidean space $R^6$ consisting of the pairs $(x,u)$ with
 $x\in X$ and $u\in R^6$. More exactly, the vector bundle in the
 present case is a continuous mapping $\pi: X\times R^6 \to X$ such
 that $\pi^{-1}(x)= \{ x \} \times R^6\cong R^6$ for all $x\in X$.}}
 $X\times R^6$ or $X\times R^4$, and coincide with each other at the
 center of mass of the observer's system. Therefore there is no
 contradiction between classical relativistic theory and quantum
 mechanical theory, even though the former is set on the curved
 Riemannian space and the latter is on the Euclidean space.

We call the Euclidean coordinates $(t,x)=(t_{n\ell},x_{n\ell})=
(t_{(H_{n\ell},\HH_{n\ell})},
x_{(H_{n\ell},\HH_{n\ell})})$ inside the local system $L=(H_{n\ell},
\HH_{n\ell})$
the {\bf proper coordinate system}, and the curved Riemannian
 coordinates associated with the observer
$L=(H_{n\ell},\HH_{n\ell})$
the {\bf observer's coordinate system}.

The curved Riemann space appears in this way only related with
 the observation, and the Euclidean space appears as the framework
 of the unobservable inner quantum mechanical world.

For the observable classical world, we assume the following Axioms 4--5.
We use the same notation
 $(t,x)=(t_{(H_{n\ell},\HH_{n\ell})},x_{(H_{n\ell},
\HH_{n\ell})})$
also to denote the classical coordinates in $X$ for the observer's system
 $(H_{n\ell},\HH_{n\ell})$, for the difference between the classical
 and quantum coordinates are only in their metric.

\F
{\bf General principle of relativity:}

\F
{\bf Axiom 4.}\ Those laws of physics which control
the {\bf relative} motions of the {\bf centers} of mass of the
{\bf observed} local systems are expressed
by the classical equations  which are covariant under the change of {\bf
observer's} coordinate systems of $R^4$:
$$
(t,x)=(t_{(H_{mk},\HH_{mk})},
x_{(H_{mk},\HH_{mk})}) \to (t,x)=(t_{(H_{n\ell},\HH_{n\ell})},x_{(H_{n\ell},
\HH_{n\ell})})
$$
for any pairs $(m,k)$, $(n,\ell)$.

It is included in this axiom that one can observe the positions of other
systems (i.e., their centers of mass) in his coordinate system $(t,x)$.

The relative velocities of the observed systems are then defined as
 quotients of the relative positions of those systems and the
(local and quantum mechanical)
time $t$ of the observer's system. These are our definitions of
the measurement procedure of {\bf classical} quantities, which  accord with
the ordinary (implicit) agreement among physicists where the time
 is given {\it a priori}.

\F
{\bf Principle of equivalence:}

\F
{\bf Axiom 5.}\
The coordinate system $(t_{(H_{n\ell},\HH_{n\ell})}, x_{(H_{n\ell},
\HH_{n\ell})})$ associated
with the local system  $(H_{n\ell},\HH_{n\ell})$ is
the  local Lorentz system of coordinates. Namely the gravitational
 potentials $g_{\mu\nu}$ for the {\bf center} of mass of the local
 system  $(H_{n\ell},
\HH_{n\ell})$, {\bf observed} in this coordinates $(t_{(H_{n\ell},
\HH_{n\ell})}, x_{(H_{n\ell},\HH_{n\ell})})$, are equal to $\eta_{\mu\nu}$.
Here $\eta_{\mu\nu}=0$ $(\mu\ne\nu)$, $=1$ $(\mu=\nu=1,2,3)$, and $=-1$
$(\mu=\nu=0)$.

We do not assume the so-called field equation which determines
 the metric $g_{\mu\nu}$. We hold the room for the equations which
 would be found preferable in the future to the present ones.

Let us remark that the difference between the proper coordinate system
 and the observer's coordinate system is their metric.
 However at the center of mass of a local system $L$,
 these coincide with each other. In fact,
at the center of mass of $L$, the Riemann metric $g_{\mu\nu}$ is equal
 to $\eta_{\mu\nu}$ by Axiom 5. Thus at any time $t$, the Riemann
 distance at the origin = the center of mass is given by
$$
d\tau^2=-g_{\mu\nu}(t,0,0,0)dx^\mu dx^\nu=-\eta_{\mu\nu}dx^\mu
 dx^\nu=dt^2-dx_1^2-dx_2^2-dx_3^2.
$$
For a person at the center of mass, $x=0$ always. Thus for him or her
$$
d\tau^2=dt^2.
$$
In this sense, the local time $t$ is identified with the relativistic
 proper time $\tau$.

The Euclidean distance inside the local system $L$ is
$$
d\ell^2=dt^2+dx_1^2+dx_2^2+dx_3^2.
$$
Thus for the center of mass,
$$
d\ell^2=dt^2
$$
again.

We set these two metrics on $X$ and on $R^4$ so that
 they coincide with each other at the center of mass =
 the origin of both coordinate systems. These metrics do not
 contradict each other, for the spaces $X$ and $R^4$
where these metrics are set are mutually orthogonal.

Summing up, we have

\F
{\bf Theorem 2.}\ Axioms 1--5 are consistent.

\section{Observation}

Given the system $(H_{mk},
\HH_{mk})$ with coordinates $(t_{mk},x_{mk})=(t_{(H_{mk},\HH_{mk})},$
$x_{(H_{mk},\HH_{mk})})$,
 we start
with the quantum mechanics, i.e., with the
Schr\" odinger propagator $e^{-it_{mk}H_{mk}}$. Then
the
quantum mechanical velocities of the particles in the system $(H_{mk},
\HH_{mk})$ are given by the quotients $q_b= x_b/t_{mk}$, asymptotically as
$t_{mk}\to\infty$,  of the position vectors $x_b$ of the particles and the
local time $t_{mk}$.

\F
{\bf Assumption on observation:}

\F
{\bf Axiom 6.}
The momenta $p_j=m_j x_j/t_{mk}$ of the particles $j$ with mass $m_j$
in the observed local system
$(H_{mk},\HH_{mk})$ with coordinate system $(t_{mk},x_{mk})$, given as above,
are observed, by the observer system $(H_{n\ell},\HH_{n\ell})$
 with coordinate
system $(t_{n\ell},x_{n\ell})$, as $p_j^\prime=m_j x_j^\prime/t_{n\ell}$,
where $x_j^\prime$ is obtained from $x_j$ by the relativistic
transformation of coordinates:  $(t_{mk},x_{mk})$ to $(t_{n\ell},x_{n\ell})$
as in Axiom 4. The same is true for the observation of the energies of
the particles: the energies of the particles in the observed local
system are observed by the observer as the ones transformed
in accordance with the relativity.

Namely it is assumed that the quantum mechanical momenta $p_j=m_j x_j/t_{mk}$
of the particles within the system $(H_{mk},\HH_{mk})$ are observed in actual
experiments by the observer system $(H_{n\ell},\HH_{n\ell})$ with coordinate
system $(t_{n\ell},x_{n\ell})$, as the {\bf classical} quantities $p_j^\prime
=m_j x_j^\prime/t_{n\ell}$ whose values are calculated
 or predicted by correcting
the quantum mechanical values $p_j$ with taking the relativistic effects of
observation into account. A similar assumption is made for the energies
of the particles.

\F
{\bf Theorem 3.}\
Axiom 6 is consistent with
Axioms 1--5.
\vskip-1mm

\F
{\it Indication of Proof:}\
 Axiom 6 is concerned only with the
quantum mechanics  {\bf within} the local system $(H_{mk},\HH_{mk})$
so that it gives the rules  to transform the {\bf quantum mechanical} values,
e.g., $p_j$, of the system $(H_{mk},\HH_{mk})$ to the {\bf classical
 mechanical} values, e.g.,
$p_j^\prime$,  that would be {\bf observed} experimentally by the observer.
 It is therefore not related with {\bf any} physics laws of the particles
{\bf within} the system $(H_{mk},\HH_{mk})$, unless the transformed values
(e.g., $p'_j$) are compared with the actual experimental values.
\vskip-1mm

In this sense, Axiom 6 is concerned only with {\bf how the nature
looks to the observer}. Together with Axioms 1--5, it gives the
 prediction of the physical
values observed in actual experiments, and is checked solely through the
experimental data.
\vskip-1mm

\section{Scattering of one neutron in a uniform gravitational field}
\vskip-1mm

\F
{\bf Experiment by Collela, Overhauser, and Werner \cite{C}:}

Consider the experiment done by Collela et al. \cite{C} of measuring
 the interference of one neutron. This experiment is described in
 some simplification as in the following Figure 1:

\newfont{\lfont}{line10}

\newcommand{\righthd}{$\vcenter{
  \hbox{\lfont\char'00}}$}

\hskip3.88cm C \hskip4.28cm  D
\vskip-4mm

\hskip4cm\righthd
\hskip-2mm---\hskip-0.4mm---\hskip-0.4mm---\hskip-0.4mm---
\hskip-2mm---\hskip-0.4mm---\hskip-0.6mm$\to$\hskip-0.6mm---
\hskip-2mm---\hskip-0.4mm---\hskip-0.4mm---\hskip-0.4mm---
\hskip-2mm---\hskip-2.4mm\righthd$\hskip-4mm\longrightarrow$
 O :Observer
\vskip-5.8mm

\hskip4.14cm $|$\hskip4.6cm $|$
\vskip-5.2mm

\hskip4.14cm $|$\hskip4.6cm $|$
\vskip-5.2mm

\hskip4.14cm $|$\hskip4.6cm $|$
\vskip-5.2mm

\hskip4.1cm $\uparrow$\hskip4.51cm $\uparrow$\hskip0.8cm the height BD $=L$
\vskip-5.2mm

\hskip4.14cm $|$\hskip4.6cm $|$
\vskip-5.2mm

\hskip4.14cm $|$\hskip4.6cm $|$
\vskip-5.2mm

\hskip4.14cm $|$\hskip4.6cm $|$
\vskip-5.8mm

\hskip3cm S $\longrightarrow$\hskip-2mm---\hskip-2mm\righthd
\hskip-2mm---\hskip-0.4mm---\hskip-0.4mm---\hskip-0.4mm---
\hskip-2mm---\hskip-0.4mm---\hskip-0.6mm$\to$\hskip-0.6mm---
\hskip-2mm---\hskip-0.4mm---\hskip-0.4mm---\hskip-0.4mm---
\hskip-2.6mm---\hskip-1.8mm\righthd
\vskip-3mm

\hskip4.26cm A \hskip4.28cm B
\vskip-4mm

\hskip6.4cm $\downarrow$
\vskip-2mm

\hskip5.8cm EARTH
\vskip-1mm

\hskip5.8cm Figure 1

\F
A neutron beam emitted at S is split into two beams by an
 interferometer at A, and the two beams are recombined at point
 D by other interferometers or mirrors B and C. The height $L$ of the line
 BD on the earth can be varied. The dependence of the relative
 phase on $L$ is given as follows,
according to the experiment of \cite{C} up to errors of about 1 \%:
\beq
\hbar^{-1}mgLT,
\ene
where $m$ is the mass of the neutron, $g$ is the acceleration of
 gravity, and $T$ is the (observed) time that the beams travel from
 C to D or A to B. This experiment shows that quantum mechanics and
 gravity play an important role {\it simultaneously} in the size of
 desktop environment. In fact, the lengths of the lines AB and BD are
 less than 10 cm in \cite{C}.

\noindent
{\bf Explanation in our theory:}

This experiment can be explained in our context, if we see it
 as a 3-body scattering phenomenon of a neutron N by two mirrors
 B and C as in Figure 2.

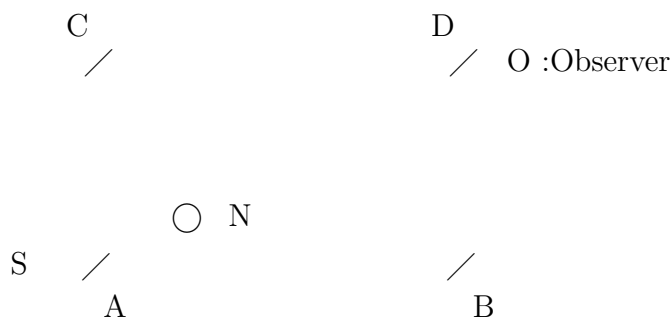
\def\MARU#1{{\rm\ooalign{\hfil\lower.168ex\hbox{#1}
\hfil\crcr\mathhexbox20D}}}

\hskip3.74cm C \hskip4.42cm  D
\vskip-4mm

\hskip4cm\righthd\hskip45mm\righthd
\hskip4mm O :Observer
\vskip-5.8mm

\hskip4.14cm $\ $
\vskip-5.2mm

\hskip4.14cm $\ $
\vskip-5.2mm

\hskip4.14cm $\ $
\vskip-5.2mm

\hskip4.14cm $\ $
\vskip-5.2mm

\hskip4.14cm $\ $
\vskip-5.2mm

\hskip5.14cm \MARU{\ } \hskip2mm N
\vskip-5.2mm

\hskip4.14cm $\ $
\vskip-5.8mm

\hskip3cm S \hskip6mm\righthd\hskip45mm\righthd
\vskip-3mm

\hskip4.24cm A \hskip4.46cm B
\vskip-4mm

\hskip6.4cm $\downarrow$
\vskip-2mm

\hskip5.8cm EARTH
\vskip-1mm

\hskip5.8cm Figure 2

\noindent
Let the masses of mirrors B, C be $M$, the neutron mass be $m$,
 and $0<m<<M$. Then the Hamiltonian of this system is
$$
H=\frac{p^2}{2m}+\frac{P_B^2}{2M}+\frac{P_C^2}{2M},
$$
where $p$, $P_B$ and $P_C$ are the momentum operators for N, B and C.
To separate the center of mass, we introduce the Jacobi coordinates
with letting
 $x$, $X_B$ and $X_C$ the coordinates of N, B and C, \beq
  x_1&=&x-X_C,\\
\nonumber
     x_2&=&X_B-\frac{mx+MX_C}{m+M},
\ene
or another Jacobi coordinates
\beq
  x_1&=&x-X_B,\\
\nonumber
  x_2&=&X_C-\frac{mx+MX_B}{m+M}.
\ene
Using these coordinates, $H$ can be written in the same form for
 both coordinates:
\beq
\nonumber
H&=&H_1+H_2, \\
\nonumber
H_1&=&\frac{p_1^2}{2\mu}, \quad   H_2=\frac{p_2^2}{2\nu}.
\ene
Here note that the variables $p_1$ and $p_2$ are mutually independent,
 hence $H_1$ commutes with $H_2$, where $p_1$ and $p_2$ are the momenta
 conjugate to $x_1$ and $x_2$, and
$\mu, \nu$ are the reduced masses:
$$
\mu^{-1}=m^{-1}+M^{-1}, \quad \nu^{-1}=M^{-1}+(m+M)^{-1}.
$$
In the following, we denote the Hamiltonians for (5.2) by
$H_1, H_2$, and for (5.3) by $H'_1, H'_2$.

We take the unit system with $\hbar=h/(2\pi)=1$. Then the propagation
 of the 3-body system is given by
\beq
\exp(-itH)f=\exp(-itH_1)\exp(-itH_2)f,
\ene
where $f(x)=f(x_1, x_2)$ is the initial wave function at the time
 $t=0$, just after the neutron has been split into two beams by the
 interferometer A.

\noindent
{\bf Remark.}\  Here the {\bf time} $t$ is the local time determined
 by the Hamiltonian $H$ or the correspondent local system, which we
 will denote by $H$ in the sequel.

The decomposition (5.4) has two forms according to the choice of
 coordinates (5.2) or (5.3). (There is another choice, but it has
 nothing to do with our argument.)

\noindent
{\bf The initial wave function $f(x_1, x_2)$:}  $x_1$ is the
 distance vector between N and C, or, between N and B, and $x_2$
 is the distance vector between B and the center of mass of the system
 N+C, or between C and the center of mass of the system N+B. Therefore,
 as seen from the formula for $x_2$ in (5.2) or (5.3), we may regard it as
$$
x_2=X_B-X_C\quad \mbox{or}\quad X_C-X_B,
$$
for $M$ is larger enough than $m$.
Thus we can regard
$x_2$ as constant during the scattering process, hence
$f(x_1, x_2)$ can be regarded as a function of $x_1$ only.

Namely $f(x_1, x_2)$ can be regarded as the wave function of the neutron
 N, and is split into two wave packets
$f_1(x_1, x_2)$, $f_2(x_1, x_2)$
 at time $t=0$ by the interferometer A:
$$
f=f_1+f_2.
$$
$f_1$ is the packet moving to the direction from A to C, and
$f_2$ is the one from A to B.

Therefore (5.4) can be rewritten as follows:
\beq
\exp(-itH)f=\exp(-itH_1)\exp(-itH_2)f_1+\exp(-itH'_1)\exp(-itH'_2)f_2.
\ene
As remarked above, we can regard $x_2=X_B-X_C$ or $X_C-X_B$, therefore
 we can think $H_2=H'_2$. Thus, noting that $H_2$ commutes with $H_1$ and
that $H'_2$ commutes with $H'_1$, we have
$$
(5.5) = \exp(-itH_2) \{\exp(-itH_1)f_1+\exp(-itH'_1)f_2\}.
$$
The description up to here is by the local time $t$ determined by
 the local system $H$.

The decomposition within $\{\ \}$ of (5.5)
gives a decomposition of the local system $H$ into two local systems
$H_1$ and $H'_1$. The reason that we could use the same local time $t$
 in these systems is that we were considering the scattering in the
 same local system $H$.

If there is no relativistic correlation between these local systems
 and the observer, the observer's time is the same as these local
 systems' times, and the observer sees the same phenomena as the ones
 with letting the observer's time $t_O$ equal to the local time $t$ of
 that local system $H$. Namely in this case, two different observations
 cannot distinguish the two local times.

However, if the gravitational field exists as in the present case,
 these local times can be distinguished by observation as follows.

$H'_1$ is the Hamiltonian consisting of N and B, and its center of
 mass is regarded as located at B by $m<<M$. Therefore that local
 system has  a lower gravitational potential in amount $gL$ compared
 to the observer O, hence the local time $t$ of the local system
 $H'_1$ is related with the observer's time $t_O$ as follows approximately:
$$
t = \frac{t_O}{\sqrt{1+(2gL)/c^2}} = t_O(1-(gL)/c^2).
$$
Therefore
$$
\exp(-itH'_1)f_2=\exp(-it_O\cdot H'_1)\exp(it_O\cdot (gL/c^2)H'_1)f_2.
$$

$H_1$ is the local system consisting of N and C, and the center of
 mass is located at C with the same height as the observer. Hence
$$
t=t_O.
$$
We note
that we can regard $H_1=H'_1$, for
 $H=H_1+H_2=H'_1+H'_2$ and $H_2=H'_2$.

{}From these, we have the following decomposition of the observational
 wave function for this 3-body system:
\beq
\exp(-itH)f=\exp(-itH_2)\exp(-it_O\cdot H_1)\{f_1+\exp(it_O\cdot gLm)f_2\}.
\ene
Here we regarded $f_1$ as the wave function of the neutron N,
 and approximated its energy by the classical energy $mc^2$.

If we calculate the asymptotic behavior as $t\to\infty$ of
 the first two factors of (5.6), we have with ${\cal F}$ being
 the Fourier transformation
\beq
\nonumber
(\exp(-itH_2)g)(x_2) &\to& C(t)\exp(itp_2^2/(2\nu))({\cal F}g)(p_2)\quad
  (p_2=x_2/t),\\
\nonumber
(\exp(-it_O\cdot H_1)g)(x_1) &\to& C(t)\exp(it_O\cdot p_1^2/(2\mu))
({\cal F}g)(p_1) \quad (p_1=x_1/t_O),
\ene
where  $C(t)$ is the constant such that the absolute value of the
 first two factors in (5.6) equals 1 asymptotically as $t\to\infty$.
 Therefore there remains only the absolute value of the parentheses
$\{\ \}$ of (5.6), which gives the desired phase difference
 and explains the interference observed in \cite{C}.

\noindent
{\bf Remarks.}

1.  In the above the neutron N is regarded as moving from  A to D
 in a classical velocity $(v_1, v_2)$.
Therefore the time necessary for N to reach D from A
is given by
$T=L/v_2$, which coincides with the time
$T=$ the length of AB$/v$, where $v$ is the horizontal velocity $=v_1$.

2. In the above scattering process, the interactions between N and
 B, C are not included in the Hamiltonian.
This point may be remedied by introducing the very short-range potentials
effective only in the vicinity of the neutron and the mirrors
 B, C. Actually these mirrors consist of a huge number of
 particles and the phenomenon should be treated as a many body problem
 including such a huge number of particles. But the above idealization
 works well for explaining the phenomenon.

\section{Hubble's law}

Hubble's law is a phenomenon that appears when one observes
 the light emitted from stars and galaxies far away from the earth.
 The emission of light itself is a quantum mechanical phenomenon that
 could be explained in the nonrelativistic quantum field theory as
 in \cite{K}, Section 11-(2). The observation or reception of this
 emission of light on the earth is explained as a classical observation
 according to our postulate Axiom 6, with assuming the
 Robertson-Walker metric as usual.

Robertson-Walker metric is the metric derived from the assumptions
 of {\it homogeneity} and {\it isotropy} of the large scale structure
 of the universe.
We refer the reader to \cite{M}, Chap. 27 for details,
 and we here only outline the argument.

Under the hypotheses of homogeneity and isotropy, the metric is
 in general of the form
$$
ds^2=-(dx^0)^2+d\sigma^2=-(dx^0)^2+a(x^0)^2\gamma_{ij}(x^k)dx^idx^j,
$$
where $x^0$ is the time parameter that `slices' the spacetime by means
 of a one parameter family of some spacelike surfaces, and
 $(x^1,x^2,x^3)$ is the `comoving, synchronous space coordinate
 system' for the universe, in the sense of \cite{M}, sections
 27.3--27.4. $a(x^0)$ is the so-called ``expansion factor"
 that describes the ratio of expansion of the universe in the
 usual context of general relativity.  A consideration by the
 use of homogeneity and isotropy
yields (\cite{M}, section 27.6) that for some functions $f(r)$
 and $h(x^0)$
$$
ds^2=-(dx^0)^2+e^{f(r)}e^{h(x^0)}\{(dx^1)^2+(dx^2)^2+(dx^3)^2\}.
$$
Assuming Einstein field equation
$G^\mu_{\ \nu}-\lambda\delta^\mu_{\ \nu}=\kappa T^\mu_{\ \nu}$
 and calculating, we get with replacing $e^{h(x^0)}$ by a constant
 times $e^{h(x^0)}$
$$
ds^2=-(dx^0)^2+e^{h(x^0)}\left(1+k\frac{r^2}{4r_0^2}\right)^{-2}(dx)^2,
$$
where $k=0$ or $+1$ or $-1$. This is called Robertson-Walker metric.
Using the polar coordinates
$(r, \theta, \varphi)$ and setting
$$
\frac{r}{r_0}=u, \quad R(t)=r_0e^{h(x^0)/2}\quad(t=x^0),
$$
 one can rewrite $ds^2$ as follows:
$$
ds^2=-(dt)^2+R(t)^2\left(1+\frac{k}{4}u^2\right)^{-2}
[du^2+u^2\{(d\theta)^2+(\sin\theta d\varphi)^2\}].
$$

Suppose $k=+1$, and consider a 3-dimensional sphere of radius
 $A$ in a 4-dimensional Euclidean space
$$
A^2=(y^4)^2+\sum_{k=1}^3(y^k)^2.
$$
The metric on this sphere is
$$
d\sigma^2=\sum_{k=1}^3(dy^k)^2+(dy^4)^2.
$$
This is rewritten by using the above equation of the sphere as follows:
$$
d\sigma^2=\sum_{k=1}^3(dy^k)^2+
\left\{A^2-\sum_{k=1}^3(y^k)^2\right\}^{-1}
\left(\sum_{\ell=1}^3y^\ell dy^\ell\right)^2.
$$
Set $\rho^2=\sum_{k=1}^3(y^k)^2$, and define $v$ by
$$
\rho=A\left(1+\frac{v^2}{4}\right)^{-1}v.
$$
Using polar coordinates
$(\rho,\theta,\varphi)$
instead of $(y^1,y^2,y^3)$,
and rewriting $\rho$ by the use of $v$, we have
$$
d\sigma^2=A^2\left(1+\frac{v^2}{4}\right)^{-2}[(dv)^2+v^2\{(d\theta)^2
+(\sin\theta d\varphi)^2\}].
$$
If we set $A=R(t)$, and identify $v$ as $u$,
this formula coincides with the space part $d\sigma^2$ of the above
 Robertson-Walker metric $ds^2$.

In this sense, the space part slice $t=$ constant of the spacetime
 can be regarded as a 3-dimensional sphere of radius $R(t)$
 in a 4-dimensional Euclidean space, hence $R(t)$ can be regarded
 as the radius of the universe.

In this context, the universe can be regarded as expanding
 when it is observed. Further the Hubble's cosmological redshift
 is  explained in this context of classical observation
also in our theory
owing to Axiom 6, as in section 29.2 of \cite{M}.

We remark that the `expansion' in this classical
 sense is different from the stationary universe $\phi$ in
 our context of quantum mechanical sense.
The former `expansion' is the result of
 an observation activity with fixing one observer's coordinate system,
 e.g., in the above explanation we have assumed a synchronous
 coordinate system,
 which explains why the universe looks expanding for all observers.
The latter quantum mechanical stationary universe $\phi$ is
 the inner structure of its own and is independent of the observer's
 coordinate system. Theorem 2 guarantees that these two views are
 consistent with each other, and Axiom 6 predicts that this framework
 would explain and resolve the problems related with the actual
 observations. In the present problem of Hubble's law and `expansion'
 of the universe, these phenomena are the consequences of the
 {\bf observation} with one coordinate system fixed. In other words,
 they are `appearance,' so to speak,  which the universe makes under
 the `interference' of the observer to try to reveal its figure or shape.
More philosophically, the past or the future does not exist unless one
 fixes the time coordinate. The `Big Bang' is an imagination under this
 {\bf assumption} of the {\it a priori} existence of time coordinate.
Unless it is observed with assuming the existence of a time coordinate,
 the universe
 is no more than a stationary state, which does not change
and is correlated within itself as a whole.

Our theory is a reflection and a clarification of this assumption of
 the existence of time,
 adopted {\bf implicitly} in almost all physics theories today.

Example of the last section is an experiment of human size, and the one
in this section is an observation of cosmological size. These two
 examples would indicate an applicability of our theory to a unified
 treatment of physical phenomena of both sizes.

\section{Uncertainty of time}

According to Derezi\'nski \cite{4}, one has for $f\in L^2(R^{3n})$
\beq
\int_1^\infty t^{-1}\left\Vert
\left|q_a-\frac{x_a}{t}\right|^{1/2}J_a\left(\frac{x}{t}\right)
h(H)e^{-itH}f\right\Vert^2 dt<\infty,
\ene
where $a$ is a cluster decomposition, $h\in C_0^\infty(R^1)$,
 and $J_a$ is a family of functions that gives a decomposition of
 configuration space.

This formula means roughly that
\beq
\nonumber
\left\Vert\left|q_a-\frac{x_a}{t}\right|^{1/2} J_a\left(\frac{x}{t}\right)
 e^{-itH}f\right\Vert &\le& C_f\quad(\mbox{uniformly\ in}\ t),\\
\nonumber
&\to& 0 \quad (\mbox{along some sequence}\ t=t_k\to\infty).
\ene
Rewriting this for the 2-body case by rereading the proof of
 Derezi\'nski, one has
\beq
\left\Vert\left(\frac{x}{t}-q\right) e^{-itH}f\right\Vert \le C_f,
\quad \to 0\ \mbox{as}\ t=t_k\to\infty.
\ene

Let us calculate the uncertainty of time by using this relation.

First let us review the usual uncertainty relation. We consider the
 2-body 1-dimensional case for simplicity.
{}From
$$
[p,x]=\frac{h}{2\pi i}I=:-iaI, \quad a\ne0,
$$
we have
$$
2{\mbox{Im}} (pf,xf)=a\Vert f\Vert^2.
$$
Then by $a\ne0$
$$
0\le \Vert f\Vert^2=\frac{2}{a}\mbox{Im}(pf,xf)\le\frac{2}{|a|}
|(pf,xf)|\le\frac{2}{|a|}\Vert pf\Vert\Vert xf\Vert.
$$
Therefore for a normalized $f$ with $\Vert f\Vert=1$
$$
\Vert pf\Vert\Vert xf\Vert\ge \frac{|a|}{2}.
$$
Set now
$$
\rho=(pf,f),\quad \sigma=(xf,f).
$$
Then applying the above calculation to $p-\rho$, $x-\sigma$,
 one has the usual uncertainty relation
$$
\Delta p\cdot\Delta x:=\Vert pf-\rho f\Vert \Vert xf-\sigma f
\Vert\ge\frac{|a|}{2}
=\frac{h}{4\pi}.
$$
If we note $p=mq$ ($m$ is the reduced mass of the present 2-body system),
 we get
\beq
\Delta x\cdot\Delta q\ge\frac{h}{4\pi m}.
\ene
(Notice that this holds even if one replaces $f$ by $e^{-itH}f$:
$$
\Delta (e^{itH}xe^{-itH})\cdot\Delta (e^{itH} qe^{-itH})\ge\frac{h}{4\pi m}.
\quad)
$$

Set
$$
R(t)=\frac{x}{t}-q.
$$
Then the above Derezi\'nski's inequality becomes
$$
\Vert R(t)e^{-itH}f\Vert\le C_f,\quad \to 0 \ \mbox{as}\ t=t_k\to\infty.
$$
As an approximate interpretation of this inequality, we take the following
$$
\left(e^{itH}\left(\frac{x}{t}-q\right)^2 e^{-itH}f,f\right)\approx0.
$$
Then this means
\beq
R(t)=\frac{x}{t}-q\approx0\quad\mbox{or}\quad \frac{x}{t}\approx q,
\ene
where we have omitted the propagators $e^{\pm itH}$ for simplicity.
Namely the expected value of time $t$ is given by
$$
t=\frac{(xe^{-itH}f,e^{-itH}f)}{(qe^{-itH}f,e^{-itH}f)}.
$$
Taking $\Delta$'s of (7.4), we have
$$
\frac{\Delta x}{\Delta t}\approx \Delta q.
$$
Therefore
$$
\Delta t\approx \frac{\Delta x}{\Delta q}=\frac{\Delta x\cdot\Delta q}
{(\Delta q)^2}.
$$
{}From the uncertainty relation (7.3), this implies
\beq
\Delta t \ge \frac{h}{4\pi m}\frac{1}{(\Delta q)^2}.
\ene
Here $(\Delta q)^2=(qe^{-itH}f,qe^{-itH}f)-(qe^{-itH}f,e^{-itH}f)^2\to0$
along $t=t_k\to\infty$. (Notice that this convergence is
very slow usually, due to (7.1).)

Conversely,
$$
\Delta t\approx\frac{(\Delta x)^2}{\Delta x\cdot\Delta q}
\le\frac{4\pi m}{h}(\Delta x)^2.
$$
Here $(\Delta x)^2=(xe^{-itH}f,xe^{-itH}f)-(xe^{-itH}f,e^{-itH}f)^2=O(t^2)$.

Thus, in the usual experiment where $t< 1$ sec, it is expected
 that the uncertainty $\Delta t$ of time is approximately of the order
\beq
c_1 \frac{h}{4\pi m}\le \Delta t\le C_2 \frac{4\pi m}{h}.
\ene
In this sense the uncertainty of time is proportional to $m^{-1}$.

For example, in the case of neutron
$$
m=167 \times 10^{-26} \mbox{grams},\quad \hbar=1.054\times 10^{-27}
 \mbox{erg}\cdot\sec
$$
yield
$$
\frac{h}{4\pi m}\approx 3\times 10^{-4} \mbox{cm}^2\cdot\sec^{-1}.
$$
The dimension of $c_1$ in (7.6) is $\sec^2\cdot\mbox{cm}^{-2}$
 by the definition of $(\Delta q)^2$ after (7.5). This implies that
 the lower bound of the left side of (7.6) is of order
$$
c_1\times 10^{-4}\sec.
$$
This is the case of 2-body system of neutron and another particle.
 (Remember that the time is not defined for one body system.)
In the actual case, we include
the macroscopic
 systems in the observed system, as the mirrors B, C in Section 5.
 Therefore the uncertainty of time
 becomes extremely low.

For example, for the system of
1 $\mu$ grams = 1$\times 10^{-6}$ grams, we have
$$
\mbox{uncertainty}\ = c_1 \times 10^{-24} \sec,
$$
and for the system of 1 gram
$$
\mbox{uncertainty}\ =c_1\times 10^{-30} \sec.
$$

\section{Concluding discussions}

We summarize the framework of our theory as in \cite{K}, Section 10:

\F
{\bf Local times:}
\begin{itemize}
\item
The times are defined only for local systems $(H_{n\ell},\HH_{n\ell})$.
\item
The
total universe $\phi$ has no time associated.
\item
The local times arise through
the affections from other particles outside the local systems. (Definitions
1--3.)
\item
The uncertainty principle holds only within these local systems as the
uncertainty of the local times.
\item
The quantum mechanics is confined within each
local system  in this sense.
\item
The quantum mechanical phenomena
between two local  systems  appear only when they are combined as
a single local system.
\item
In the local system,
the interaction and forces propagate
with infinite velocity or in other words they are {\bf unobservable}.
\end{itemize}

\F
{\bf Local systems:}

\begin{itemize}
\item
Each local system can be the observer of other systems.
\item
In this situation the
local systems are {\bf mutually independent} in the sense that
the associated quantum  mechanical local times are not correlated
in general.
\item
Therefore there are no reasons to exclude the classical
mechanics in describing the {\bf observable}
relative behavior of the observed systems with respect to the observer.
\item
Thus
the gravitational potentials can be introduced
  in accordance with the theory of general relativity.
\item
These potentials determine the global space-time
structure around the observer system.
\item
Inside the observer system
the space-time  is  Euclidean.
\item
The observer itself cannot detect the
 gravitational correlation
 or  the space-time structure inside its own system.
\item
On the contrary, between the
local systems, the observer can detect
 only the classical mechanical effects.
\item
Nevertheless, through the media
(e.g., light in classical sense) which connect the observer and
the observed systems and obey the classical physics,  the observer sees,
through some relativistic corrections of the observed classical values,
that the physics laws inside the other local systems follow the quantum
 mechanics.
\end{itemize}

\F
{\bf Total universe:}

\begin{itemize}
\item
These facts are all the consequences of the introduction of
 {\bf local times} which are proper to each local system.
\item
The time is neither
a given thing nor a common one to the total universe.
\item
On the contrary
there can be defined no global time. More strongly the total universe is a
(stationary) bound state of the total Hamiltonian $H$ of infinite degrees
of freedom.
\item
The times arise only
when the observers restrict their attention to its subsystems as
approximations of the total Hamiltonian $H$.
\item
The universe itself
is correlated within it as a bound state of $H$.
\item
The observer
 always separates a subsystem
from it, so to speak, artificially, and the
 (steady) motion and time appear.

\item
Inside the subsystem this local time explains the quantum effects,
and outside the subsystem it explains the gravitation and
 the classical mechanics. The
relativistic quantum phenomena are explained as the relativistic effects of
the observation of the non-relativistic quantum systems.
\item
All these physical
phenomena occur by this artificial separation of the universe.
 The universe itself does not \lq change': It is a stationary bound state.
\end{itemize}

\end{document}